# Reply to Comment on "Failure of the work-Hamiltonian connection for free-energy calculations" by Horowitz and Jarzynski

The Comment [arXiv:0808.1224] by Horowitz and Jarzynski (HJ) obtains as a main result a general free energy change for a harmonic system that in the macroscopic limit does not recover the textbook expression for the energy change of a Hookean spring. The reason is that HJ improperly identify work with parametric changes of the Hamiltonian instead of with the standard quantity, force times displacement.

Consider a macroscopic Hookean spring with elastic constant $k$ that is stretched by an applied external force that increases slowly from $0$ to $f_{ext}$. As explained in textbooks [1, 2], the work performed by the external force on the spring is $W = f_{ext}^2 / 2k$ and consequently the energy stored in the spring increases by

$$\Delta E = W = f_{ext}^2 / 2k \ . \qquad (1)$$

This basic example does not require the use of Hamiltonian or statistical mechanics descriptions and provides unequivocal results accepted since the establishment of Newton's laws of motion.

HJ obtain, however, a decrease in energy, $\Delta E_Z = -f_{ext}^2 / 2k$, independent of the temperature and applicable to any system size, which therefore should have been valid in the macroscopic limit, when fluctuations can be neglected. Why is HJ's result not in accordance with Newton's laws of motion?

The connection of HJ's result with classical mechanics fails because parameters (e.g. an external force), in general, are not generalized coordinates, in contrast to what HJ assume, and consequently parametric changes of the Hamiltonian cannot be identified with force times displacement, namely with work. As explained in Ref. [3], the general Hamiltonian describing the stretching of the spring along the coordinate $x$ by the external force $f_{ext}$ is $H(x; f_{ext}) = \frac{1}{2} k x^2 - f_{ext}(x - \gamma)$, where $\gamma$ is an arbitrary parameter that does not affect the system in any way. Yet, $\gamma$ determines the quantity $\int (\partial H / \partial f_{ext}) \dot{f}_{ext} dt = f_{ext} \gamma - \frac{1}{2k} f_{ext}^2$ identified by HJ with free energy changes. HJ, in their Eq. (3), arbitrarily choose $\gamma = 0$.

HJ have shown that $\Delta S = 0$ and therefore that $\Delta G = \Delta E$. A well-established result of statistical mechanics is that the partition function gives the free energy. Why does HJ's statistical mechanics approach using the partition function not recover the macroscopic limit, when thermal fluctuations are not relevant?

The connection with thermodynamics fails not because the partition function does not give the free energy but because the reference point of the energy in HJ's calculations is not the same in states A and B, as implied by the fact that changes in the Hamiltonian are not the work performed on the system. As a trivial illustration of this point, consider the Hamiltonian $H(x;g) = H_0(x) + g$, with $g$ a time-dependent parameter. Changes of $g$ would lead to arbitrary energy changes $\Delta E_Z = g_B - g_A$ when in fact the system remains unaltered.

In the Comment, HJ also argue that free energy changes are negative for spontaneous processes but can have any sign for nonspontaneous processes. This argument is incompatible with the fact that the free energy is a state function. Indeed, Eq. (3) of HJ's Comment implies a positive change in free energy ($G_A - G_B = f_{ext}^2 / 2k$) for the spontaneous relaxation of a spring upon removal of the external force, which contradicts the thermodynamic criterion for spontaneity.

In thermodynamics, free energy changes are always defined as the work, $W = Force \times Displacement$, performed over a reversible trajectory [4, 5]. For small systems, the work is averaged because it fluctuates [6]. For instance, the quantity that has been used experimentally to compute free energy changes of RNA molecules is the average of $W$ over reversible trajectories [7], not parametric changes of the Hamiltonian as HJ propose.

If the standard definition of work is used [3], none of the inconsistencies observed in the free energy changes computed by Horowitz and Jarzynski are present. In particular, the standard definition of work leads to well-defined free energy changes that depend only on physical parameters, are negative for spontaneous processes, are positive for nonspontaneous processes, and recover the textbook expression for the energy of a Hookean spring [3].

J. M. G. Vilar[1] and J. M. Rubi[2]

[1]Computational Biology Program, Memorial Sloan-Kettering Cancer Center, New York, New York 10065, USA

[2]Departament de Fisica Fonamental, Universitat de Barcelona, Diagonal 647, 08028 Barcelona, Spain